\documentclass[10pt, conference, letterpaper]{IEEEtran}
\usepackage[pdftex]{graphicx}
\usepackage{graphicx}
\usepackage{subfigure}
\usepackage[justification=centering,font=scriptsize]{caption}
\usepackage{amsfonts}
\usepackage{amsmath}

\usepackage{amsthm}
\usepackage{array}
\usepackage[ruled,linesnumbered]{algorithm2e}
\SetAlCapNameFnt{\footnotesize}
\SetAlCapFnt{\footnotesize}
\usepackage[english]{babel}
\usepackage{booktabs}
\usepackage[pdftex,colorlinks,linkcolor=black,citecolor=black,urlcolor=black]{hyperref}
\usepackage{multirow}
\newtheorem{thm}{Theorem}

\ifCLASSINFOpdf
\else
\fi

\begin{document}
\title{Crowdsourced Live Streaming over the Cloud}

\author{\IEEEauthorblockN{Fei Chen$^{*\dag}$,\ Cong Zhang$^{*}$,\ Feng Wang$^{\ddag}$,\ Jiangchuan Liu$^{*}$ }
\IEEEauthorblockA{$^{*}$School of Computing Science, Simon Fraser University, Canada\\
$^{\dag}$School of Digital Media, Jiangnan University, China\\
$^{\ddag}$Department of Computer and Information Science, The University of Mississippi, USA\\
Email: chenf@jiangnan.edu.cn, congz@sfu.ca, fwang@cs.olemiss.edu, jcliu@cs.sfu.ca}}

\maketitle

\begin{abstract}
%\boldmath

Empowered by today's rich tools for media generation and distribution, and the convenient Internet access, {\em crowdsourced streaming} generalizes the
single-source streaming paradigm by including massive contributors for a video channel. It calls a joint optimization along the path from crowdsourcers, through streaming servers, to the end-users to minimize the overall latency. The dynamics of the video sources, together with the globalized request demands and the high computation demand from each sourcer, make crowdsourced live streaming challenging even with powerful support from modern cloud computing. In this paper, we present a generic framework that facilitates a cost-effective cloud service for crowdsourced live streaming. Through adaptively leasing, the cloud servers can be provisioned in a fine granularity to accommodate geo-distributed video crowdsourcers. We present an optimal solution to deal with service migration among cloud instances of diverse lease prices. It also addresses the location impact to the streaming quality. To understand the performance of the proposed strategies in the realworld, we have built a prototype system running over the planetlab and the Amazon/Microsoft Cloud. Our extensive experiments demonstrate that the effectiveness of our solution in terms of deployment cost and streaming quality.

\end{abstract}


% \fi

\section{Introduction}
The Internet has witnessed a significant increase in the popularity of media streaming with multi-source channels. In traditional video broadcast,
the content of a channel generally comes from a single source, though it could be replicated and then streamed from different servers in a Content Distribution Network (CDN).
A \textit{multi-source} system, however, not only serves massive audience worldwide, but its content also comes from multiple contributing sources. For example, since Feb. 17, 2012, NASA Television's Public and Media channels began to transmit their respective content in high definition (HD), with live feeds from such space centers as the NASA Headquarters, the Johnson Space Center, and the Goddard Space Flight Center \footnote{\url{http://www.nasa.gov/multimedia/index.html}}. With their respective content sources, they collectively serve the users interested in the stories and the latest news from NASA. In the very recent 2014 Sochi Winter Olympics, NBC had a total of 41 live feeds distributed both in Solchi and in the USA \footnote{\url{http://www.istreamplanet.com/sochi-2014/}}, and in FIFA World Cup 2014, when a goal was scored, CBC synchronized the live scenes of the cheering fans in the public squares from a number of cities worldwide in its live streaming channel. The evolution is driven further by today's advanced mobile/tablet devices that can readily capture high quality video anywhere and anytime (e.g. iPhone 6 supports both 60 fps 1080p video recording, and 720 fps slow-motion recording for 720p videos), and such mainstream video sharing platforms as YouTube and Veedme have already enabled multi-party collaborative video content production. All these together are shifting the video service paradigm from the conventional single source, to multi-source, to many source, and now toward {\em crowd source}, in which the available video sources for the content of interest become highly diverse and scalable.

%\begin{figure}
%\centering
%% Use the relevant command to insert your figure file.
%% For example, with the graphicx package use
%  \includegraphics[width=1.0\linewidth]{map.pdf}
%\caption{NASA Tv Live Broadcast among Different Centers}
%\label{fig:0}
%      % Give a unique label
%\end{figure}

Global streaming imposes high demand on end device capabilities and network connections. The situation is further complicated in a crowdsourced streaming system. First,  crowdsourced videos are geo-distributed: they come from all over the world, and then spread all over the world. Not only the scale of the consumers is enormous, but also is that of the contributors; Second, the crowdsourcers are often much more dynamic than dedicated content providers, as they can start or terminate a video contribution as their own will. This is particularly true when non-professionals using their smartphones/tablets for video production; Third, for collective content production, massive server capacity is necessary to deal with online video synchronization, processing, and transcoding for highly heterogeneous video contributors and consumers. For example, Twitch TV, the world's leading video platform and community for gamers \footnote{\url{http://www.twitch.tv/}}, allows any of its users to broadcast their live streaming videos online through their PCs or PS3/XBOX consoles.  It attracts over 44 million visitors per month, and every second its servers are loaded with thousands of live channels. For such a large system, significant effort is needed to collect the highly dynamic and distributed video streams online, and to process and distribute the live channels to subscribers all over the world.

Elastic resource provisioning and computation offloading are where \textit{cloud computing} platforms excel \cite{scaling}. We have seen many new generation of cloud-based multimedia services
that emerged in recent years, e.g., Netflix, which are rapidly changing the operation and business models in the market. Facing similar scale challenges, crowdsourced live streaming would benefit from the cloud services, too. Yet the distributed and highly dynamic sources, as well as the much more stringent delay constraints imposed by live streaming, make the problem more involving, which remains to be explored with novel and distinct solutions.

In this paper, using realworld measurement, we identify the potential benefits as well as the key challenges when crowdsourced video meets cloud. We present a generic framework for a cost-effective cloud service that provisions cloud resources in a fine granularity to work with geo-distributed video crowdsourcers. Using adaptive and collaborative leasing strategies, our design well accommodates the diverse capacities and prices of cloud instances, and addresses the location impact to the streaming quality. We have built a prototype system running over the Internet and the Amazon EC2/Microsoft Azure cloud, and the experimental results experiments demonstrate the effectiveness of our solution in terms of both deployment cost and streaming quality.

The remainder of this paper proceeds as follows. Section 2 discusses the background and related work. Section 3 presents an overview of the crowdsourced live streaming system, and analyzes its unique challenges using realworld data traces. In Section 4, we first investigate the inherent problem of cloud leasing strategy. An optimal solution is then developed to deal with geo-distributed crowdsourcers in Section 5. In Section 6, we present a prototype platform with the measurement results and the trace-driven simulation. Finally, Section 7 concludes the paper and discusses potential future directions.

\section{Background and Related Work}
In the past two decades, video streaming over the Internet has quickly risen to become a mainstream "killer" application \cite{decade}. For large scale distribution, many existing systems rely on content distribution networks (CDNs) \cite{unreeling} or peer-to-peer (P2P)~\cite{Opportunities}, or hybrid solutions \cite{CALMS}. More recently, with the flexible and elastic resource provisioning, cloud computing has been proven to be an efficient solution toward highly scalable video distribution. A prominent example is Netflix,\index{Netflix} a major on-demand Internet video provider. Netflix migrated its entire infrastructure to the powerful Amazon AWS cloud in 2012, using EC2 for transcoding master video copies to 50 different versions for heterogeneous end users and S3 for content storage~\cite{unreeling}. In total, Netflix has over 1 petabyte of media data stored in
Amazon's cloud. It leases the computation, bandwidth and storage
resources with much lower long-term costs than those
with over-provisioned self-owned servers, and reacts better and faster to user
demand with the dramatically increasing scale. There have been pioneer studies on migrating video services to the cloud to accommodate worldwide-distributed and time-varying video demands~\cite{scaling} \cite{decade}. Aggarwal et al. \cite{IPTV} showed that the cost of IPTV services can be noticeably reduced through a cloud infrastructure, and Wu et al. \cite{scaling} utilized a geo-distributed cloud to support large scale social media streaming applications. Wang et al. \cite{CALMS} presented CALMS (Cloud-Assisted Live Media Streaming) to lease and adjust cloud server resources in a fine granularity, meeting with the temporal and spatial dynamics of demands from online users.

%A game theory based model was also developed to maximize the resource utilization in datacenters for video streaming~\cite{bargain}.

Empowered by today's rich tools for media generation and collaborative production, and the convenient Internet access, {\em crowdsourcing} further extends the
single-source paradigm. It combines the efforts of multiple self-identified contributors, known as {\em crowdsourcers}, for a greater result, and has seen success in many areas~\cite{mobile}. For example, LiFS (Locating in Fingerprint Space) was developed for wireless indoor localization with smartphones based crowdsourcing \cite{indoor}. Ou et al. \cite{Signal} used crowdsourcing approach to optimize mobile devices' energy efficiency by utilizing signal strength traces shared by other devices in cellular networks. For video applications, a scalable system that allows users to perform content-based searches on continuous collection of crowdsourced video was proposed in~\cite{crowdvideo}. Biel et al. \cite{Impressions} investigated the the crowdsourcing of personal and social traits in online social video or social media content in general. Recently, Youtube has integrated with Google Moderator, a crowdsourcing and feedback production, to increase the engagement between viewers and content creators.  Such other video sharing sites as Poptent and VeedMe  have also opened interfaces for crowdsourcers with user generated content. Crowdsourced live streaming services have emerged in the market as well, especially for streaming sports online broadcast. Examples include Stream2Watch.me and sportLEMON.tv.

Our study is motivated by these pioneer works. Yet crowdsourced live streaming demands efficient content collection, processing, and distribution with stringent delay constraints, which remain to be explored. This
paper highlights these unique challenges, particularly when crowdsourced live streaming meets cloud, and presents our initial attempts toward addressing these challenges.

\begin{figure}
\centering
  \includegraphics[width=0.8\linewidth]{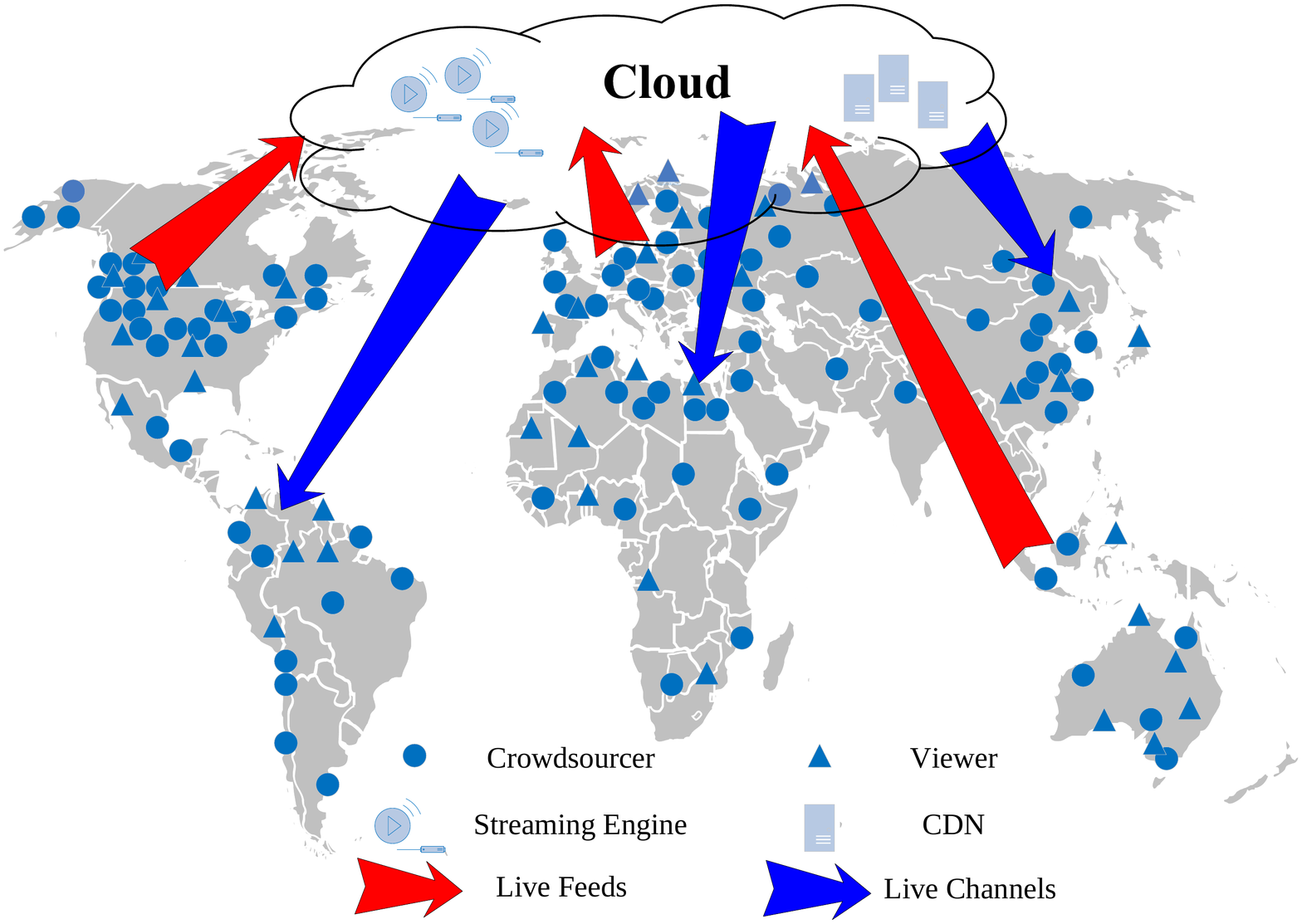}
\caption{A generic crowdsourced live streaming system over cloud}
\label{fig:1}
\vspace{-0.5cm}       % Give a unique label
\end{figure}

\section{Crowdsourced Living Streaming: System Overview and Challenges}

\begin{figure*}[htb]
\begin{minipage}[t]{0.33\linewidth}
\centering
\includegraphics[width=1\textwidth]{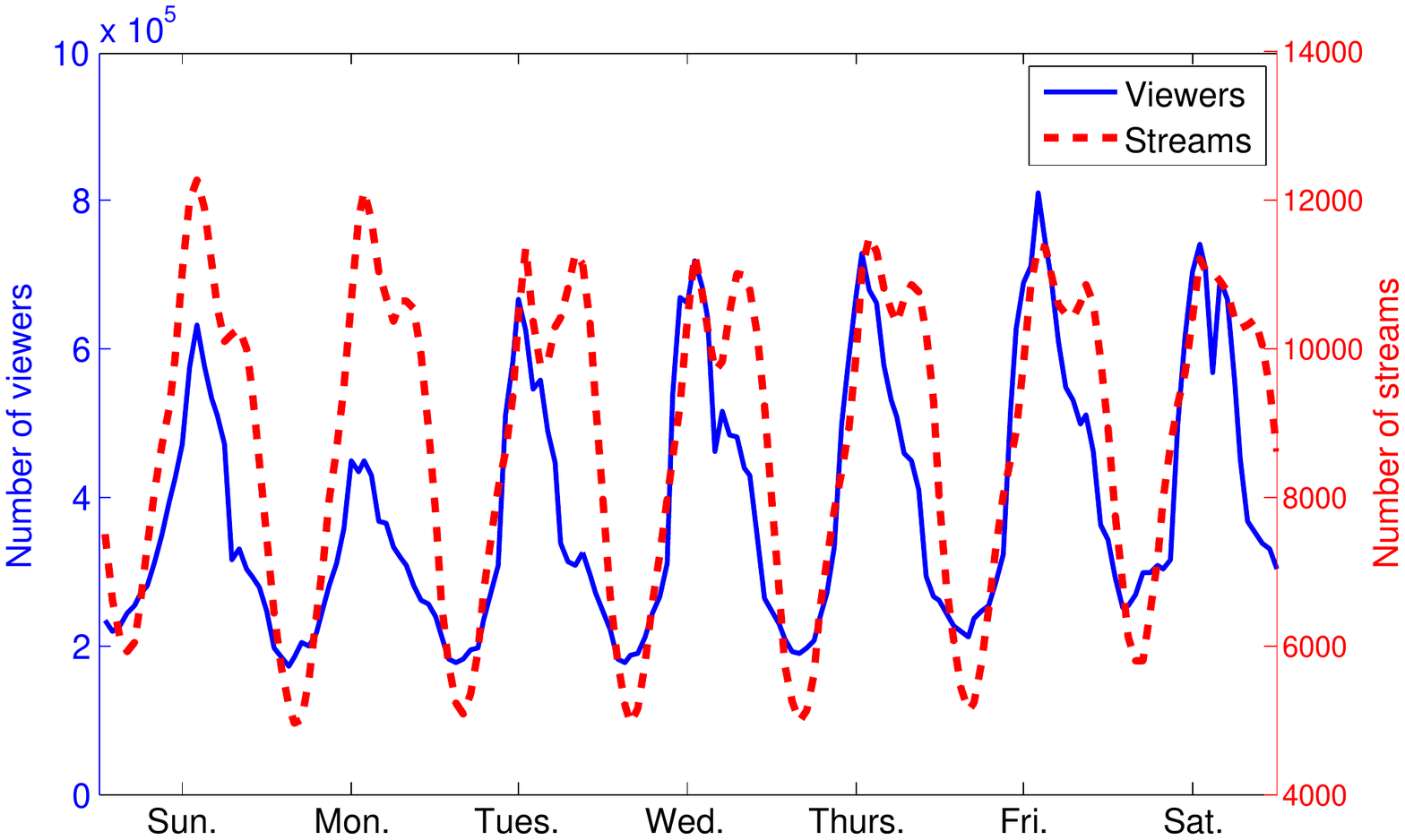}
\caption{Number of viewers and source streams
variation in one week}
\label{week}
\end{minipage}
\begin{minipage}[t]{0.33\linewidth}
\centering
\includegraphics[width=1\textwidth]{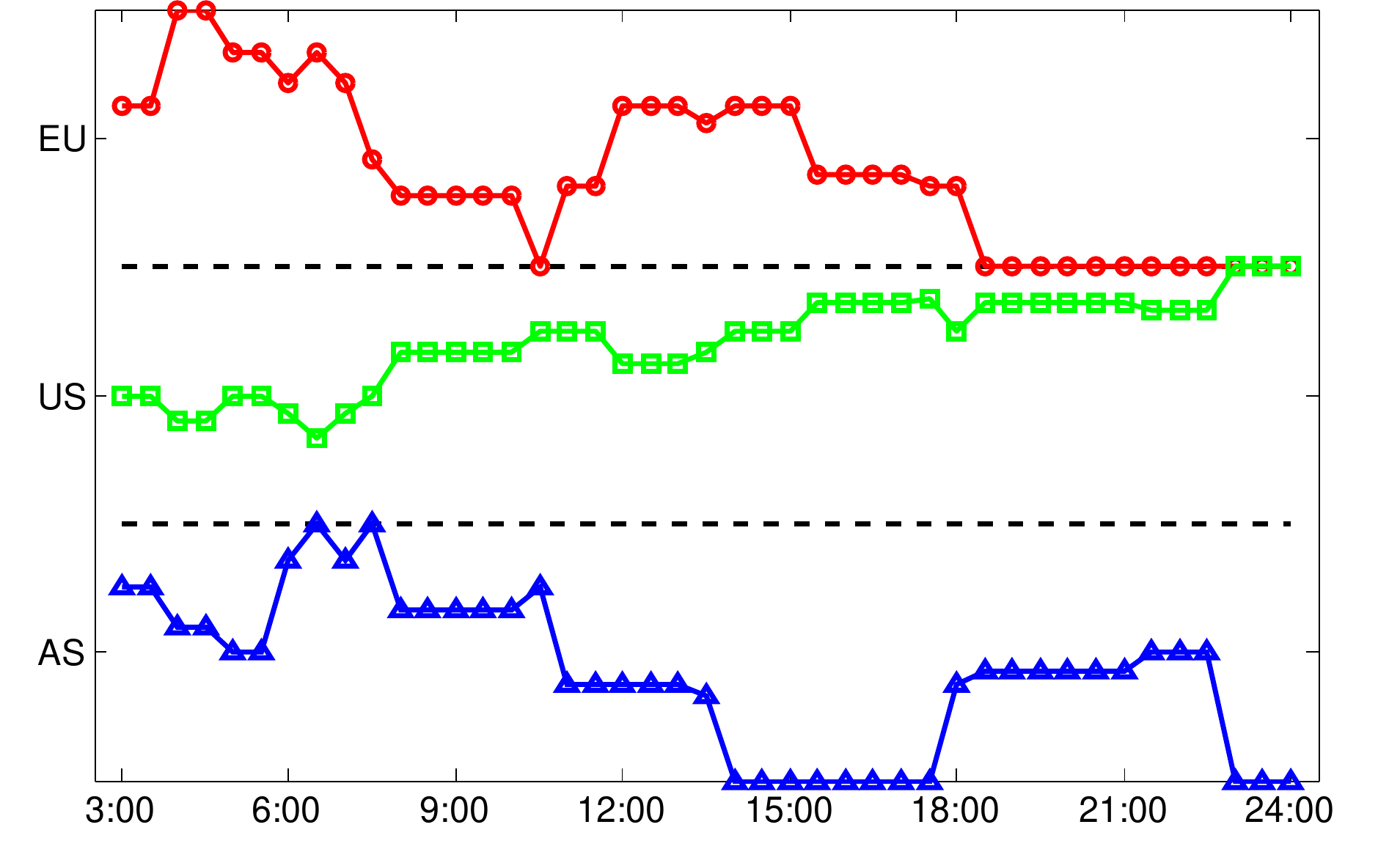}
\caption{Source stream distribution in one day}
\label{number}
\end{minipage}
\begin{minipage}[t]{0.33\linewidth}
\centering
\includegraphics[width=1\textwidth]{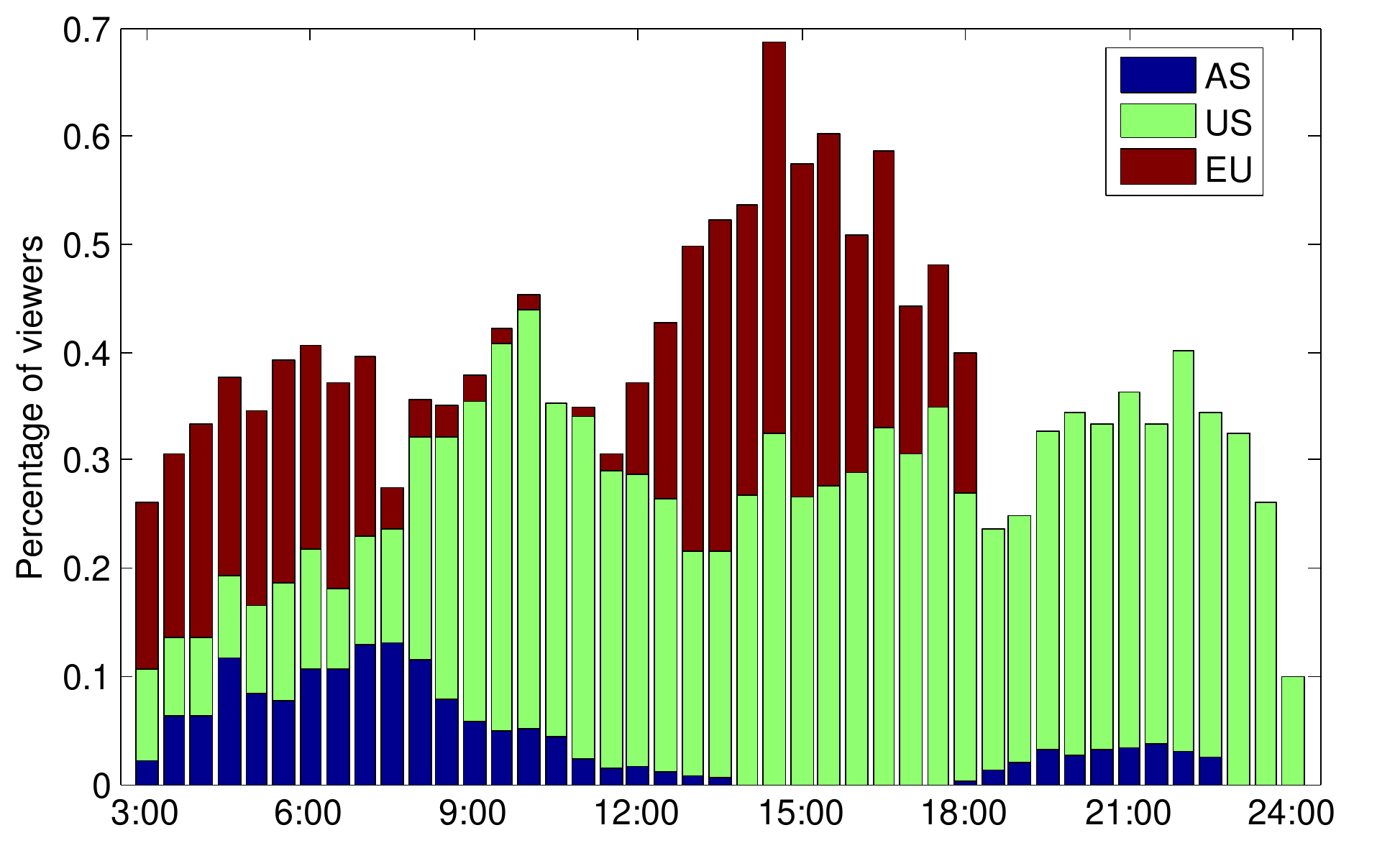}
\caption{Viewer demand for the distributed source streams in one day}
\label{viewer}
\end{minipage}
\end{figure*}

We illustrate a generic crowdsourced live streaming system with geo-distributed {\em crowdsourcers} and {\em viewers} in Fig. \ref{fig:1}. A set of crowdsourcers (or {\em sourcers} in short) upload their individual video contents in realtime, which, through a video production engine, collectively produce a single video stream. The stream is then lively distributed to viewers of interest. Both the sourcers and viewers can be heterogenous, in terms of their network bandwidth, and their hardware/software configurations for video capture and playback. As such, realtime transcoding is necessary during both uploading and downloading, so as to unify the diverse video bitrates/formats from different sourcers for content production, and to replicate the output video stream to serve the heterogeneous viewers, possibly through  through a CDN with such adaptation mechanisms as DASH (Dynamic Adaptive Streaming over HTTP)~\cite{Adaptive}.

This generic architecture reflects that of state-of-the-art realworld systems. For example, NBC's video content from the 41 feeds in Sochi Winter Olympics were encoded by Windows Azure Media Services to the 1080P format, and dynamically transcoded into HLS and HDS formats. These streams were then pulled from Azure to the Akamai's CDN and distributed to audiences on targeted devices, resulting in over 3000 hours of live Olympics streaming contents.

Given the large system scale and the high bandwidth, storage, and computation demands involved, cloud services with elastic resource provisioning is expected. We again consider a generic geo-distributed cloud infrastructure, which consists of multiple \textit{cloud sites} distributed in different geographical locations (e.g., US East (N. Virginia) and EU (Ireland) in Amazon EC2 Cloud)\cite{scaling}. Each cloud site resides in a data center, and contains a collection of interconnected and virtualized servers. The server resources will be provisioned for crowdsourced live streaming, e.g., computation resources for collective production and transcoding.

Optimization for conventional single-source video streaming is generally {\em viewer-driven}; the resource provisioning depends on the distribution of the viewers. In crowdsourced video, however, the sourcers themselves come from all over the world, whose distribution must be as well taken into account during resource provisioning. This is further aggravated given that the collaborative production escalates the demands on both bandwidth and computation. The crowdsourced streaming workflow is also much more dynamic, as individual sourcers can start/terminate based on their own schedules.

\begin{table}
\centering
\caption {Top 5 sourcers from Twitch.tv on July, 12th}
\begin{tabular}{ccc}
\toprule[1.5pt]
Sourcers ID & Time (Pacific Time) & Location  \\
\hline
$riotgames$ & 11:10 AM-15:40 PM & Cologne, Germany  \\
\hline
$dota2ti\_ru$ & 7:10 AM-18:10 PM & Seattle, USA  \\
\hline
$srkevo1$ & 6:00 AM-23:40 PM & Las Vegas, USA  \\
\hline
$riotgamesturkish$ & 1:30 AM-7:10 AM & Istanbul, Turkey  \\
\hline
\multirow{2}*{$ongamenet$} & 3:00 AM-13:30 PM & \multirow{2}*{Seoul, South Korea}  \\
                         & 18:20 PM-22:40 PM &  \\

\bottomrule[1.5pt]
\end{tabular}
\label{tab:number}
\end{table}

To better understand the inherent challenges of deploying such a system, we have crawled one-week trace from July 6 to July 12, 2014 in Twitch.tv website, which has 14 geo-distributed ingest servers, 1 from Asia area (AS for short), 6 from European area ( EU for short), and 7 from United States area (US for short) to broadcast live game streams to viewers in a global scale. For simplicity, we consider that one live stream is contributed by only one sourcer. Fig. \ref{week} shows the number variation of viewers and streams in a week, from July 6 to July 12, 2014. First, it is obvious that the number of viewer is highly dynamic, which is prevalent in current large scale systems \cite{decade}. Due to the differences in time zones and languages, the distribution of viewers can be time-varying, which has been discussed in previous works \cite{Joint-wang} \cite{CALMS}. Similar to the number of viewers, we can see that the number of source streams also has great time variations in one-day time, from about 5000 streams in the early morning to almost 12000 streams in the afternoon. To further investigate the time-varying distribution of the source streams, we have measured the top 15 streams with the highest viewer population from 3:00 AM to 24:00 PM (PST) on July 12, 2014, and list the five most popular streams in Table 1. We can see that not only the time periods but also the locations of the stream sourcers are highly dynamic. In Fig. \ref{number}, we divide the locations as AS, EU, and US, and record the percentage of source streams from each region for every 30 minutes between 3:00 AM to 24:00 PM. It can be easily observed that most of the streams from Asia and Europe are during the morning and afternoon, and the number of streams from the United States keeps growing when night falls. We further measure the viewer population for the distributed source streams from each region in Fig. \ref{viewer}. We can see that in the early morning between 3:00 AM and 7:00 AM, most of the popular streams come from Europe or Asia. We conjecture that it is because the local times in Europe or Asia are in afternoon or evening, and there are more online sourcers from these regions during that time. Meanwhile, the viewer demand from these areas can also be more active during this period. And most of the viewers may prefer the streams with native language speaking sourcers. Similar reasons can also explain the increase of viewer demand for the source streams from the United States after 15:00 PM.

In summary, in a crowdsourced live streaming system, both the number and the distribution of the crowdsourcers can be highly dynamic. Together with time-varying viewer demand, the conventional server allocation design faces more challenging in a large scale. We will utilize the cloud service to coordinate the crowdsourcers and viewers. The cloud server instances (e.g. EC2 in Amazon Cloud) are provisioned to collect and process the live feeds of the crowdsourcers, and the cloud CDNs (e.g. CloudFront in Amazon Cloud) are deployed to handle the viewer dynamics. Through dynamic cloud leasing, we will present a cost-effective solution with streaming quality guarantee.

\section{Cloud-Assistance for Crowdsourced Live Streaming}
In this section, we first model the global cloud service leasing strategy with quality guarantee, and transform it into an equivalent problem in a directed graph. We will then present an optimal algorithm and an efficient online heuristic solution based on the equivalent problem.

\subsection{Problem Formulation}

We use $\mathbb{A}$ to denote the global areas, which can be divided into $n$ different regions as $\mathbb{A}=\{A_{1}, A_{2},..., A_{n}\}$. Assume that there are $m$ cloud sites all over the world, represented as $\mathbb{S}=\{s_{1}, s_{2}, ..., s_{m}\}$. As most cloud providers have a minimum unit time for the duration of leasing a server (e.g. 1 hour for Amazon EC2), we use $T$ to denote this duration. We define a {\it time slice} as an integer multiple $\kappa$ ($\kappa \in \mathbb{N}^{+}$) of $T$ and at the beginning of each time slice $\kappa T$, our cloud leasing strategy makes decisions on whether to provision or terminate the cloud servers in the distributed regions. We assume that the schedules of crowdsourced streams are predictable and can be known beforehand, where the rationale is of two folds. First, in practice a large portion of crowdsourced streams are driven by well-scheduled events (e.g. as one of the top 5 sourcers from Twitch.tv in Table~\ref{tab:number}, the channel of $srkevo1$ has a strict schedule about the Evolution 2014 Tournament\footnote{\url{http://evo2014.s3.amazonaws.com/brackets/index.html}}). Moreover, many self-motivated crowdsourcers prefer a regular broadcast schedule everyday to attract more viewers. We can accordingly forecast the numbers and distributions of both crowdsourcers and viewers for the next time slice, e.g., using techniques from \cite{CALMS}\cite{forecast}.

For a given time $t$, we denote the set of source streams from the crowdsourcers as $\mathbb{L}(t)$. According to the location distribution of crowdsourcers, we can specify the set as $\mathbb{L}^{A}(t)=\{l_{A_{1}}(t), l_{A_{2}}(t), ..., l_{A_{n}}(t)\}$ for the $n$ different regions, respectively. As all these live streams are served by the provisioned cloud instances, we further consider the set according to the dedicated cloud sites as $\mathbb{L}^{s}(t)=\{l_{s_{1}}(t), l_{s_{2}}(t),..., l_{s_{m}}(t) \}$, where $l_{s_{j}}(t)$ represents the live streaming sources loaded in cloud site $s_{j}$. For example, if $l_{s_{j}}(t)=\emptyset$, no crowdsourced stream is served by cloud site $s_{j}$, i.e., cloud site $s_{j}$ does not need to be leased at time $t$. Otherwise, if the live streams from area $A_{2}$, $A_{3}$, and $A_{5}$ are served by cloud site $s_{j}$, we have $l_{s_{j}}(t)=l_{A_{2}}(t) \cup l_{A_{3}}(t) \cup l_{A_{5}}(t)$.
%
%\begin{equation*}
%    l_{s_{j}}=\bigcup\limits_{l_{A_{i}} \in \{A_{2}, A_{3}, A_{5}\}}l_{A_{i}}
%\end{equation*}

We denote the server provisioning cost at time $t$ as $C^{p}(t)=\sum\nolimits_{j=1}^{m} c_{j}^{p}(l_{s_{j}}(t))$, where $c_{j}^{p}$ is the price of the leased instances in cloud site $s_{j}$. We assume that there is always a bootstrapping server $s_{0}$ redirecting the global live sources to the distributed streaming servers with the cost $c_{0}$. To offload the bandwidth support for the diverse viewer demands from the cloud servers, a globalized CDN strategy (e.g., CloudFront in Amazon) is deployed to distribute the live streams all over the world. The cost of out-bound traffic from the cloud servers to the CDN can be calculated by the number of channels loaded in the cloud servers, and denoted as $C^{b}= \sum\nolimits_{j=1}^{m} c_{j}^{b}(l_{s_{j}}(t))$. As the cost of the bandwidth support from the CDN to the global viewers is proportional to the viewer demands $D(t)=\sum\nolimits_{i=1}^{n} D_{l_{A_{i}}}(t)$, where $D_{l_{A_{i}}}(t)$ represents the viewer demands for the crowdsourced streams from region $A_{i}$, we can denote the total cost of the CDN as $C^{d} = c^{d}(D(t))$ with $c^{d}$ as the cost to support one unit of the viewer demand. The total cost of the crowdsourced live streaming system can thus be calculated as follows:
\vspace{-0.5cm}

%\begin{equation*}
%    \mathbb{C}=\mathbb{C}^{s}+\mathbb{C}^{b}=\{ c_{1}^{s}(l_{A_{1}})+c_{1}^{b}(l_{A_{1}}), c_{2}^{s}(l_{A_{2}})+c_{2}^{b}(l_{A_{2}}), ..., c_{m}^{s}(l_{A_{m}})+c_{m}^{b}(l_{A_{m}}) \}=\{ c_{1}(l_{A_{1}}), c_{2}(l_{A_{2}}), ..., c_{m}(l_{A_{m}}) \}
%\end{equation*}
\begin{eqnarray*}
Cost_{total} &=& c_{0}+ C^{p} + C^{b} + C^{d} \\
&=& c_{0}+ \sum\limits_{j=1}^{m} \left[c_{j}^{p}(l_{s_{j}}(t))+c_{j}^{b}(l_{s_{j}}(t))\right] + c^{d}(D(t)) \\
&=& c_{0}+ \underbrace{\sum\limits_{j=1}^{m} c_{j}(l_{s_{j}}(t))}_{Cost_{lease}} + c^{d}(D(t))
\end{eqnarray*}
where $c_{j}(\cdot)$ can be determined by the price policy of instance leasing and data traffic in cloud site $s_{j}$. As the first and last costs on the right side of the equation can not be reduced, we focus on minimizing the middle part of the total cost, i.e., the cloud leasing cost, which we denote as $Cost_{lease}$.

We assume that the live crowdsourcers in each region $l_{A_{i}}(t)$ have a preference value on a given cloud site $s_{j}$, which we denote as $P(l_{A_{i}}(t), s_{j})$. Generally, the preference value can be quantified according to the RTT, jitter or packet loss values of the connections between the crowdsourcers and the given cloud site, such as defined as a concave decreasing function of the estimated latency or a concave increasing function of the estimated connection speed in a geo-distributed service~\cite{Joint-wang}. To guarantee the streaming quality of the crowdsourced streams in region $A_{i}$, we only consider allocating these streams to the cloud sites with the top $k$ preference values, and define the set of these cloud sites as $Index(l_{A_{i}}(t), k)$ for the crowdsourced streams $l_{A_{i}}(t)$. As a real world example, Twitch/Justin.tv provides an ingest server ranker program to feedback the list with top 3 servers for each crowdsourcer.

The cloud service leasing problem in our geo-distributed crowdsourced live streaming system can thus be formulated as to find a cloud site leasing strategy $\mathbb{L}^{s}$, subjecting to the following constraints:\\

\noindent (1) Cloud site service constraint:

\begin{equation*}
    \forall A_{i} \in \mathbb{A},  \; \exists l_{s_{j}} \in \mathbb{L}^{s}, \; l_{A_{i}} \subseteq l_{s_{j}}
\end{equation*}

\vspace{-0.6cm}
\begin{equation*}
    \forall l_{s_{j}},\ l_{s_{\hat{j}}} \in \mathbb{L}^{s}, \text{if} \; j \neq \hat{j}, \ l_{s_{j}} \cap l_{s_{\hat{j}}}=\emptyset
\end{equation*}

\noindent (2) Crowdsourcer preference constraint:

\begin{eqnarray*}
  & \forall A_{i} \in \mathbb{A} , \; s_{j} \in \mathbb{S}, \;  \text{if} \; l_{A_{i}}\subseteq l_{s_{j}} \\
  & {s_{j}} \ \in \  Index(l_{A_{i}}(t), k)
\end{eqnarray*}

\noindent (3) Total budget constraint:

\begin{equation*}
    Cost_{lease} + c_{0} + C^{d} \leq Cost_{max}
\end{equation*}

The cloud site service constraint states that the crowdsourced live streams in a given region are served by only one cloud site. The preference constraint guarantees that the crowdsourced live streams in each region are collected by one of the cloud sites with the corresponding top $k$ preference values. The total budget constraint demands that the total cost including the bootstrapping server, the provisioned cloud sites and the CDN utilization must not exceed the total budget $Cost_{max}$. Our objective is to maximize the \textit{global relative preference} of the crowdsourcers, which is defined as:
\begin{equation*}
\begin{split}
  P_{global} =& \frac{\sum\limits_{\forall s_{j} \in \mathbb{S}, \ l_{A_{i}}\subseteq l_{s_{j}}}  |D_{l_{A_{i}}}(t)| \cdot P(l_{A_{i}}(t), s_{j})
  }{\sum\limits_{\forall A_{i} \in \mathbb{A}} |D_{l_{A_{i}}}(t)| P(l_{A_{i}}(t),Index(l_{A_{i}}(t), 1))}\\
\end{split}
\end{equation*}
where for ease of exposition, we also use $Index(l_{A_{i}}(t), 1)$ to denote the top $1$ preferred cloud site for the live crowdsourced streams $l_{A_{i}}(t)$. We use $|D_{l_{A_{i}}}(t)|$ to represent the size of viewer demands for crowdsourced streams $l_{A_{i}}(t)$, and $P_{global}$ is thus a relative ratio ranged between (0, 1] in the global scale.

%\begin{equation*}
%    Cost_{lease}= \sum\limits_{j=1}^{m} c_{j}(l_{s_{j}}(t))
%\end{equation*}

\begin{figure}
\centering
  \includegraphics[width=1.0\linewidth]{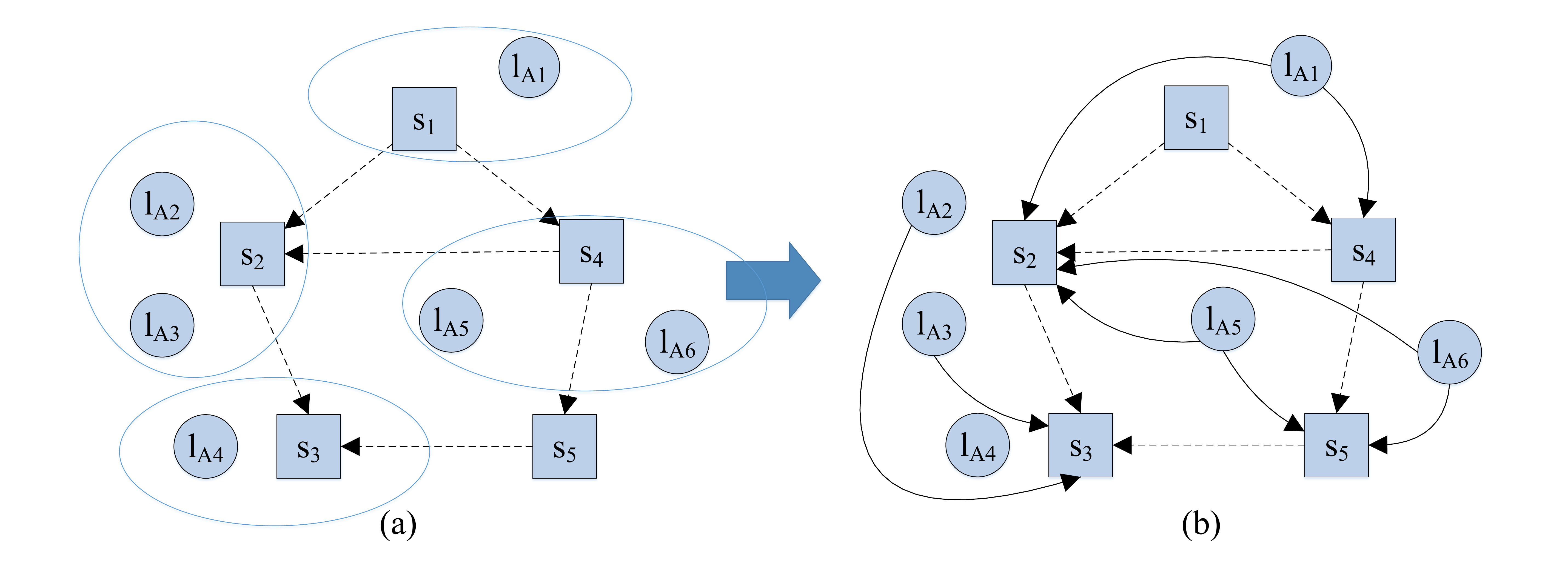}
\caption{An illustrative example of (a) distribution graph; (b) service migration vectors}
\label{fig:exmaple}
\vspace{-0.3cm}     % Give a unique label
\end{figure}

To make our solution cost-effective, we also need a second objective, i.e., to minimize the cloud leasing cost $Cost_{lease}$. It is easy to see that these two objectives (i.e., $P_{global}$ and $Cost_{lease}$) may contradict with each other, since always leasing the top preferred cloud server can increase the leasing cost. Therefore, we adopt the following linear combination form to align them together by different weights:

%\begin{equation*}
%    \forall q_{A_{i}}(s_{A_{i}}, s_{j}) \quad \text{for} \quad A_{i} \in A \; \text{and} \; s_{j} \in \mathbb{S},   \quad q_{i}(s_{A_{i}}, s_{j}) \ \leq \  \theta_{s}
%\end{equation*}

\begin{equation*}
     \frac{p \cdot Cost_{lease}}{Cost_{max}-c_{0}-C^{d}} + q \cdot (1 - P_{global})
\end{equation*}
where $p$ and $q$ are two parameters that can assign different weights to the two goals. As $P_{global}$ is a relative ratio of the preference values of all the crowdsourcers in the system (i.e. if $P_{global}=1$, all the crowdsourced live streams are allocated in their most preferred cloud sites), $(1-P_{global})$ should be minimized as $Cost_{lease}$. To make the leasing cost part also be a ratio ranged between (0, 1], we further divide $Cost_{lease}$ by $(Cost_{max}-c_{0}-C^{d})$ and then use parameters $p$ and $q$ to linearly combine the two parts together. In the next subsection, we will transform this problem to an equivalent graph problem and then propose an optimal solution.

\subsection{Equivalent Problem}

For ease of exposition, we assume the given time is $t$ for the remainder of this section and thus omit $(t)$ in all such notations as $l_{A_{i}}(t)$, $l_{s_{j}}(t)$, $D_{l_{A_i}}(t)$, etc.
Given the geo-distributed crowdsourcers and cloud sites, we can construct a \textit{distribution graph}. Fig.~\ref{fig:exmaple}(a) shows an example of 5 cloud sites and global crowdsourcers located in 6 regions. There are two types of vertices in the distribution graph, namely, the live crowdsourcers (e.g. $l_{A_{1}},...,l_{A_{6}}$ in Fig.~\ref{fig:exmaple}(a)), which are represented by circles, and the cloud sites (e.g. $s_{1},...,s_{5}$), which are represented by squares. Initially, all the live source steams are attached to their most preferred cloud sites and we denote the corresponding leasing cost as
\begin{equation*}
Cost_{initial}=\displaystyle\sum_{\forall A_{i} \in \mathbb{A}, \ s_j = Index(l_{A_{i}},1)}c_{j}(l_{A_{i}})
\end{equation*}

According to the price strategy $c_{j}(\cdot)$ of different cloud site $s_j$, we have the \textit{direction edges} between these distributed cloud sties. We use $\vec{d}(i, j)$ to denote a direction edge from the cloud site $i$ with higher price to the cloud site $j$ with lower price (e.g. in Fig. \ref{fig:exmaple} (a), $\vec{d}(4, 2)$ means that $c_{2}(x) < c_{4}(x)$ for the same crowdsourcer $x$), which indicates that the service is migrating towards a more cost-effective solution.

\begin{figure}
\centering
  \includegraphics[width=1.0\linewidth]{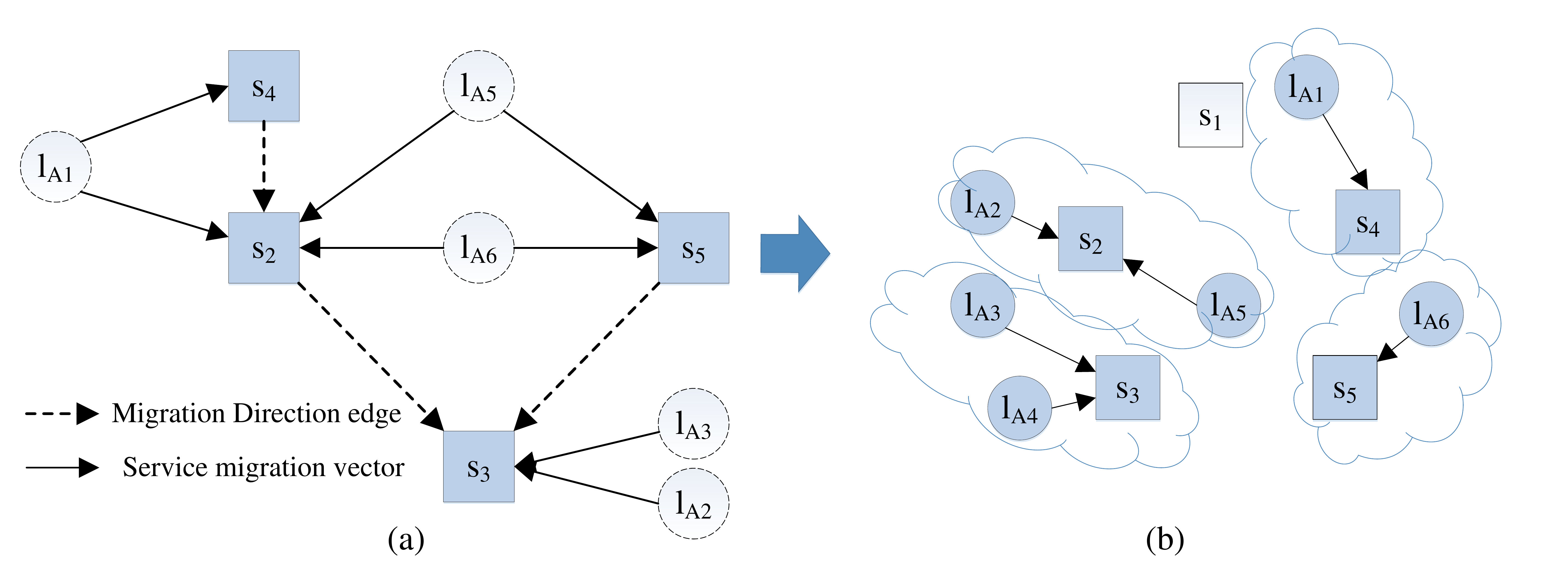}
\caption{An illustrative example of (a) a constructed service migration graph; (b) migrated cloud service for geo-distributed crowdsourcers}
\label{fig:process}
\vspace{-0.3cm}      % Give a unique label
\end{figure}

With the distribution graph and direction edges, we then generate \textit{service migration vectors} to indicate the available cloud sites for more cost-effective service migration. We use $\vec{m}(i, j)$ to denote a service migration vector that represents the live crowdsourcers $l_{A_{i}}$ are migrated and served by the cloud site $s_{j}$, rather than the cloud site $Index(l_{A_{i}}, 1)$. For example, in Fig. \ref{fig:exmaple}(b), the cloud site $s_{4}$ is preferred by the live crowdsourcers $l_{A_{5}}$ and $l_{A_{6}}$, i.e., $s_{4}=Index(l_{A_{i}}, 1)$ for $i \in \{5,\ 6\}$. According to the direction edges $\vec{d}(4, 2)$ and $\vec{d}(4, 5)$, we can have the service migration vectors $\vec{m}(5, 2)$ and $\vec{m}(5, 5)$ for the live crowdsourcers $l_{A_{5}}$, and $\vec{m}(6, 2)$ and $\vec{m}(6, 5)$ for the live crowdsourcers $l_{A_{6}}$. Define $M$ as the set of all service migration vectors that are generated from the given distribution graph. For each service migration vector $\vec{m}(i, j) \in M$, the \textit{relative preference degradation} for live crowdsources $l_{A_{i}}$ to be served by the cloud site $j$ can be calculated as follows:
\begin{equation*}
    \emph{Deg}(i, j)=\frac{|D_{l_{A_{i}}}|(P(l_{A_{i}}, Index(l_{A_{i}}, 1))-P(l_{A_{i}}, s_{j}))}{\sum\limits_{\forall A_i \in \mathbb{A}} |D_{l_{A_{i}}}| P(l_{A_{i}},Index(l_{A_{i}}, 1))}
\end{equation*}
Also, for each $\vec{m}(i, j)$, we have the \textit{cost saving} as follows:

\begin{equation*}
    Save(i, j)=c_{\hat{j}}(l_{A_{i}})-c_{j}(l_{A_{i}})
\end{equation*}
where $c_{\hat{j}}$ is the pricing policy of cloud site $s_{\hat{j}}=Index(l_{A_i},1)$.

Traversing all the service migration vectors $\vec{m}(i, j) \in M$, we can have a \textit{service migration graph} $G(V, E)$. Fig.~\ref{fig:process}(a) shows an example of Fig.~\ref{fig:exmaple}(b). We connect the cloud sites with at least one service migration vector through migration direction edges. Note that there may be more than one migration direction edges leaving from the same cloud sites. For example, in Fig.~\ref{fig:exmaple}(b) there are two migration direction edges $\vec{d}(4, 2)$ and $\vec{d}(4, 5)$ leaving from cloud site $s_{4}$. Since the set of service migration vectors $M$ has already been generated from the migration direction edges, we can put any one of these directed edges into the constructed service migration graph (which is only for the connectivity purpose that will be further explained in the next subsection). Finally, we connect the crowdsourcers to the cloud sites by the service migration vectors. In the constructed service migration graph $G(V, E)$, we can further define the \textit{optimal service migration (OSM)} problem as to find a set of migration vectors $O \subseteq M$, subjecting to the following constraints:\\

\noindent (1) Service migration vector constraint:

%\begin{equation*}
%    \forall i, \; j \in \{1,...,n\} \ \text{if} \ \exists \vec{m}(i,j) \neq 0 \ \text{then} \forall k \in \{1,...,n\} \ k \neq j \vec{m}(i,j)=0
%\end{equation*}
%
\begin{eqnarray*}
  & \forall \vec{m}(i, j), \vec{m}(i, \hat{j}) \in M \text{ and } j \neq \hat{j} \text{, }\\
  &  \text{if } \vec{m}(i,j) \in O \text{, then } \vec{m}(i, \hat{j})\notin O
\end{eqnarray*}

\noindent (2) Preference degradation constraint:

\begin{equation*}
    \forall \vec{m}(i, j) \in O, \; s_{j} \in Index(l_{A_{i}}, k)
\end{equation*}

\noindent (3) Cost saving constraint:

\begin{equation*}
\begin{split}
Cost_{initial}- \sum\limits_{ \forall \vec{m}(i, j) \in O} Save(i,j) +c_{0}+ C^{d} \leq Cost_{max}
\end{split}
\end{equation*}
The service migration vector constraint represents that there is at most one migration vector leaving from a live crowdsourcer vertex, which corresponds to the cloud site service constraint in the cloud leasing problem. The preference degradation constraint is related to the crowdsourcer preference constraint of the cloud leasing problem. The cost saving constraint refers to the total cost not exceeding $Cost_{max}$ in the original problem. Our objective is to minimize the linear combination of cost saving and the relative preference degradation as follows:

%\begin{equation*}
%   \sum\limits_{\forall \vec{m}(i, j) \in O}  \frac{p \cdot (Cost_{initial}-Save(i,j))}{Cost_{max}-c_{0}-C^{d}} + q \cdot \emph{R\_deg}(i, j)
%\end{equation*}

\begin{eqnarray*}
& & \frac{p}{Lease_{max}} (Cost_{initial} - \sum\limits_{\forall \vec{m}(i, j) \in O} Save(i, j)) \\
& & + q \big (1-(1-\sum\limits_{\forall \vec{m}(i, j) \in O} Deg(i, j)) \big ) \\
&=& \frac{p \cdot Cost_{initial}}{Lease_{max}} +
 \sum\limits_{\forall \vec{m}(i, j) \in O} \big ( q \cdot Deg(i, j)\\
 & & - \frac{p \cdot Save(i, j)}{Lease_{max}} \big )
\end{eqnarray*}
where $Lease_{max}=Cost_{max}-c_{0}-C^{d}$. As $Cost_{initial}$ cannot be further reduced, our objective can thus be simplified as to minimize
\begin{equation*}
\sum\limits_{\forall \vec{m}(i, j) \in O} \big ( q \cdot Deg(i, j) - \frac{p \cdot Save(i, j)}{Lease_{max}} \big )
\end{equation*}
%\begin{figure}
%\centering
%% Use the relevant command to insert your figure file.
%% For example, with the graphicx package use
%  \includegraphics[width=1.0\linewidth]{shortest.pdf}
%\caption{An Illustration of a Minimal Cost Forest}
%\label{fig:3}       % Give a unique label
%\end{figure}

The \textit{OSM} problem in graph $G(V, E)$ can be naturally related to the cloud site leasing problem: the optimal solution $O$ indicates the service allocation for the crowdsourcers in each region toward the distributed cloud sites. Fig.~\ref{fig:process}(b) shows an example with $O =\{\vec{m}(1, 4), \vec{m}(3, 3), \vec{m}(5, 2), \vec{m}(6, 5) \}$. Therefore, we have the set of live crowdsourcers served in each cloud site as follows: $l_{s_{1}}=\emptyset$, $l_{s_{2}}=l_{A_{2}} \bigcup l_{A_{5}}$, $l_{s_{3}}=l_{A_{3}} \bigcup l_{A_{4}}$, $l_{s_{4}}=l_{A_{1}}$, and $l_{s_{5}}=l_{A_{6}}$.

\section{Optimal Cloud Leasing Strategy}

%\begin{figure}
%\centering
%% Use the relevant command to insert your figure file.
%% For example, with the graphicx package use
%  \includegraphics[width=1.0\linewidth]{tree.pdf}
%\caption{An illustration of a minimal cost spanning forest}
%\label{fig:tree}      % Give a unique label
%\end{figure}

The optimal solution of the equivalent problem can be computed according to the spanning trees in the service migration graph. Clearly, a spanning tree is a subgraph of the directed graph $G(V, E)$. Let $T$ denote the number of spanning trees in a service migration graph $G(V, E)$. We define the set of service migration vectors in the $i$-th spanning tree ($i \in \{1,...,T\}$) as $M_{i}$, and the optimal solution of $M_i$ as $O_{i}$. We then have the following theorem:

%\begin{figure}
%\centering
%% Use the relevant command to insert your figure file.
%% For example, with the graphicx package use
%  \includegraphics[width=1.0\linewidth]{tree.pdf}
%\caption{An illustration of enumerated spanning trees}
%\label{fig:tree}      % Give a unique label
%\end{figure}

\begin{thm}\label{thm1}
There must exist an optimal solution $O$ of the service migration vectors $M$ in the service migration graph $G(V, E)$, such that $O \in \{O_{1},...,O_{T}\}$.
\end{thm}

We can prove this using contradiction by assuming that there exits an optimal solution set of the service migration vectors $O$ with edges in a circle. Then there are two scenarios if the edges in directed graph contain a circle:(1) The directed edges are sequenced in a line one after another, with the end vertex sending toward the head vertex. (2) There is more than one directed edge leaving from the same vertex. As there is no edge sending toward to the live crowdsourcers vertex in directed graph $G(V, E)$, there would be cloud sites sequenced in a circle, and we have the confliction $ c_{1}(l)>c_{2}(l)>...>c_{end}(l)>c_{1}(l)$. Also we can eliminate the scenario 2 according to the definition of service migration graph. Due to space limitation, here we omit the details of the proof, which can be found in our technical report presented.

According to Theorem~\ref{thm1}, each spanning tree can provide a local optimal solution, and the global optimal solution can be achieved by exploring all the spanning trees in $G(V, E)$. There are extensive studies on enumerating all the spanning trees in a directed graph~\cite{finding}\cite{Enumerating}. E.g., a well-known algorithm in~\cite{finding} uses backtracking and a method for detecting bridges based on the depth-first search with the time complexity $O(|V|+|E|+|E|\cdot|T|)$ and the space complexity $O(|V|+|E|)$. For a spanning tree $i$ in the service migration graph $G(V, E)$, the service migration vectors $M_i$ (and each of its subsets) are feasible solutions under the service migration vector constraint. By enforcing the preference degradation constraint, a number of spanning trees can be further screened out. Thus, for a remained spanning tree $i$, we need to calculate the local optimal migration vector set $O_{i}$ to minimize the combinational objective with the cost saving constraint, which can be solved by the classic 0-1 knapsack problem. In particular, let $\mathbb{F}(ItemSet, TotalWeight )$ denote the standard 0-1 knapsack problem. The $ItemSet$ is $M_{i}$ in our problem and the $TotalWeight$ is equal to $(\sum\limits_{\vec{m}(i, j) \in M_{\lambda}} Save(i,j)-Save_{min})$, where $Save_{min} = Cost_{initial} +c_{0}+ C^{d} - Cost_{max}$. We thus need to select a set of items $\bar{M}$ (service migration vectors) in the $ItemSet$ ($M_i$) with the total weight $\displaystyle \sum_{\forall \vec{m}(i, j) \in \bar{M}}Save(i,j) \leq \sum\limits_{\vec{m}(i, j) \in M_{\lambda}} Save(i,j)-Save_{min}$ so as to maximize the total value

\begin{equation*}
\sum\limits_{\forall \vec{m}(i, j) \in \bar{M}} \big ( q \cdot Deg(i, j) - \frac{p \cdot Save(i, j)}{Lease_{max}} \big )
\end{equation*}
From the optimal solution $\bar{O}$ of $\mathbb{F}$, we can thus calculate the optimal solution $O_i$ of $M_i$ on the spanning tree $i$ as $O_i = M_i - \bar{O}$. Then the global optimal solution can be found through enumerating all the spanning trees on the service migration graph $G(V, E)$. We summarize this optimal solution in Algorithm~\ref{dp}.

\begin{algorithm}[t]\tiny
$O=\emptyset$ \\
\For {\emph{each enumerated spanning tree} $\lambda$ \emph{on} $G(V, E)$}
{
    \If {\emph{tree} $\lambda$ \emph{fulfils the preference degradation constraint}} {
    \If {$\sum\limits_{\vec{m}(i, j) \in M_{\lambda}} Save(i,j) \geq Save_{min}$}
    {
        $\bar{O}=\mathbb{F}( M_{\lambda}, \sum\limits_{\vec{m}(i, j) \in M_{\lambda}} Save(i,j)-Save_{min} )$ \\
       $ O_{\lambda}= M_{\lambda} - \bar{O}$ \\
       \If {$objective(O_{\lambda})<objective(O)$}
       {
           $O=O_{\lambda}$
       }
    }}
}
\textbf{return} $O$ as the global optimal solution for $G(V, E)$
\caption{Optimal service migration()}
\label{dp}
\end{algorithm}
\textfloatsep=2pt

It is worth noting that finding the optimal solution for the standard 0-1 knapsack problem can become a time-consuming task as the crwodsourcers are distributed in a large scale, which can cause the optimal solution proposed in Algorithm~\ref{dp} less suitable in practice, especially for an online system with highly dynamic crowdsourcer distribution and viewer demand. To this end, we further propose a simplified heuristic algorithm in Algorithm~\ref{gd}, which can work efficiently and still return the global optimal solution under certain situations. We then have the following theorem:

\begin{algorithm}[!htp]\tiny
$O=\emptyset$ \\
\For {{\emph{each enumerated spanning tree} $\lambda$ \emph{on} $G(V, E)$}}
{
  \If {\emph{tree} $\lambda$ \emph{fulfils the preference degradation constraint}} {
    $O_{\lambda}=\emptyset$\\
    $Total_{save}=0$ \\
    sort $\vec{m}(i, j) \in M_{\lambda}$ with $\frac{Deg(i, j)}{Save(i, j)}$ in increasing order \\
    \For {$\vec{m}(i, j) \in M_{\lambda}$}
    {
        \If{($q \cdot Deg(i, j) < \frac{p }{Lease_{max}} \cdot Save(i, j) )$ \emph{\textbf{or}} $ (Total_{save} < Save_{min}$)}
        {
            put $\vec{m}(i, j)$ into $O_{\lambda}$
            $Total_{save} = Total_{save} +Save(i, j)$
        }
    }
    \If {$objective(O_{\lambda})<objective(O)$}
    {
        $O=O_{\lambda}$
    }
  }
}
\textbf{return} $O$ as the online solution for graph $G(V, E)$
\caption{Efficient online service migration()\label{gd}}
\end{algorithm}

\begin{thm}\label{thm2}
Algorithm \ref{gd} can return the global optimal solution when $Cost_{initial} + c_{0} + C^{d} \leq Cost_{max}$ for each enumerated spanning tree.
\end{thm}

Note that, if we can prove that the local optimal solution in each spanning tree can be achieved by Algorithm~\ref{gd} when $Cost_{initial} + c_{0} + C^{d} \leq Cost_{max}$, we can then prove that Algorithm~\ref{gd} can return the global optimal solution by Theorem~\ref{thm1}. We can prove this using contradiction by assuming that there is a spanning tree $\lambda$ with $Cost_{initial} + c_{0} + C^{d} \leq Cost_{max}$ but has an optimal solution $\acute{O}_{\lambda} \subseteq M_{\lambda}$, which is better than the solution $O_{\lambda}$ found by Algorithm~\ref{gd}. As $Save_{min}=Cost_{initial} + c_{0} + C^{d}-Cost_{max} \leq 0$, we always have $Total_{save} \geq Save_{min}$. Thus, for all $\vec{m}(i, j) \in O_{\lambda}$, we have $q \cdot Deg(i, j) < \frac{p }{Lease_{max}} \cdot Save(i, j)$. The contradiction can thus be achieved by first identifying the difference between $\acute{O}_{\lambda}$ and $O_{\lambda}$, and then showing that making changes to $\acute{O}_{\lambda}$ according to $O_{\lambda}$ can further improve $\acute{O}_{\lambda}$. Due to space limitation, here we omit the details of the proof, which can be found in our technical report presented.

\begin{figure*}[htb]
\begin{minipage}[t]{0.34\linewidth}
\centering
\includegraphics[width=1\linewidth]{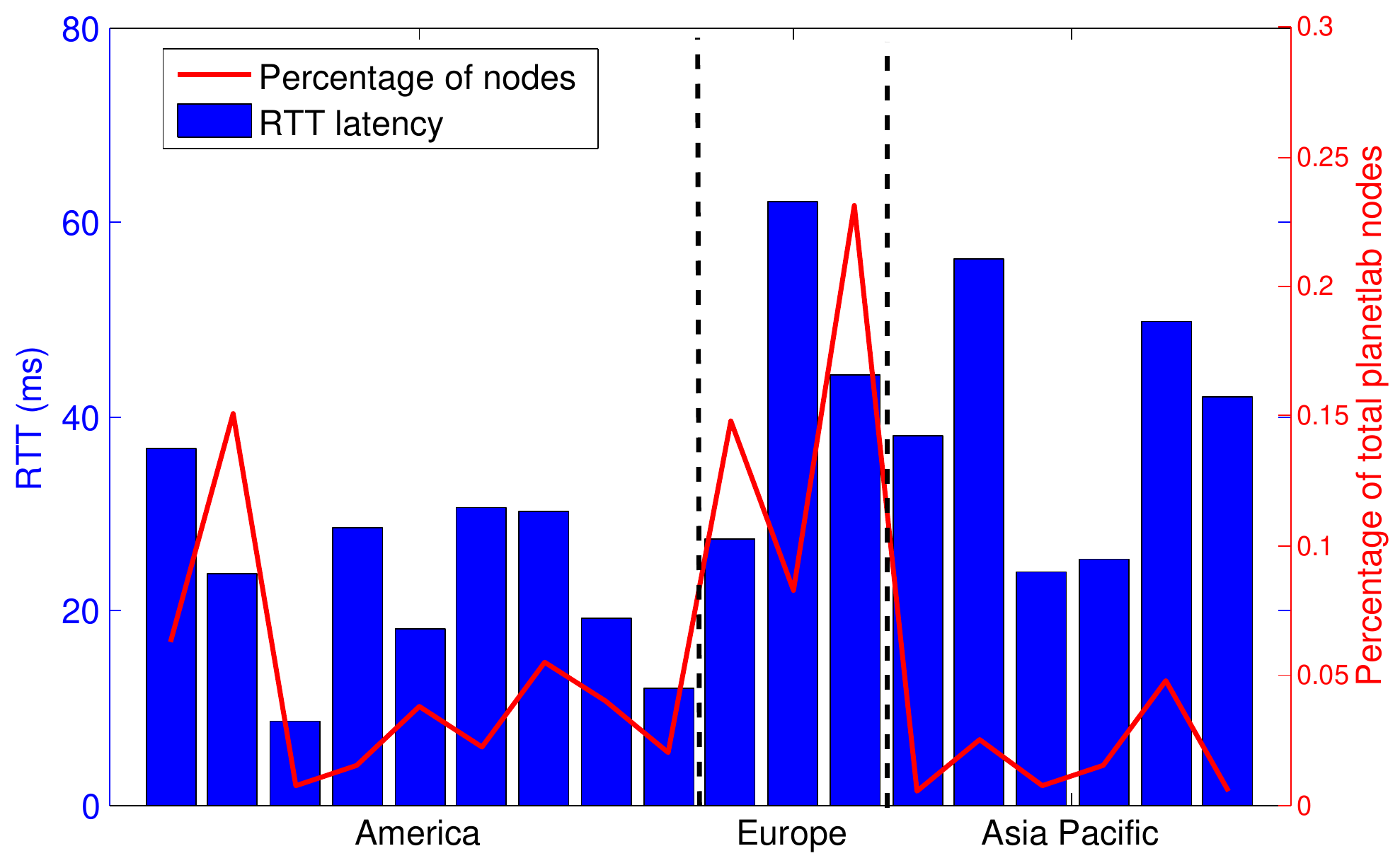}
\caption{RTT latency}
\label{fig:rtt}
\end{minipage}
\begin{minipage}[t]{0.32\linewidth}
\centering
\includegraphics[width=1\linewidth]{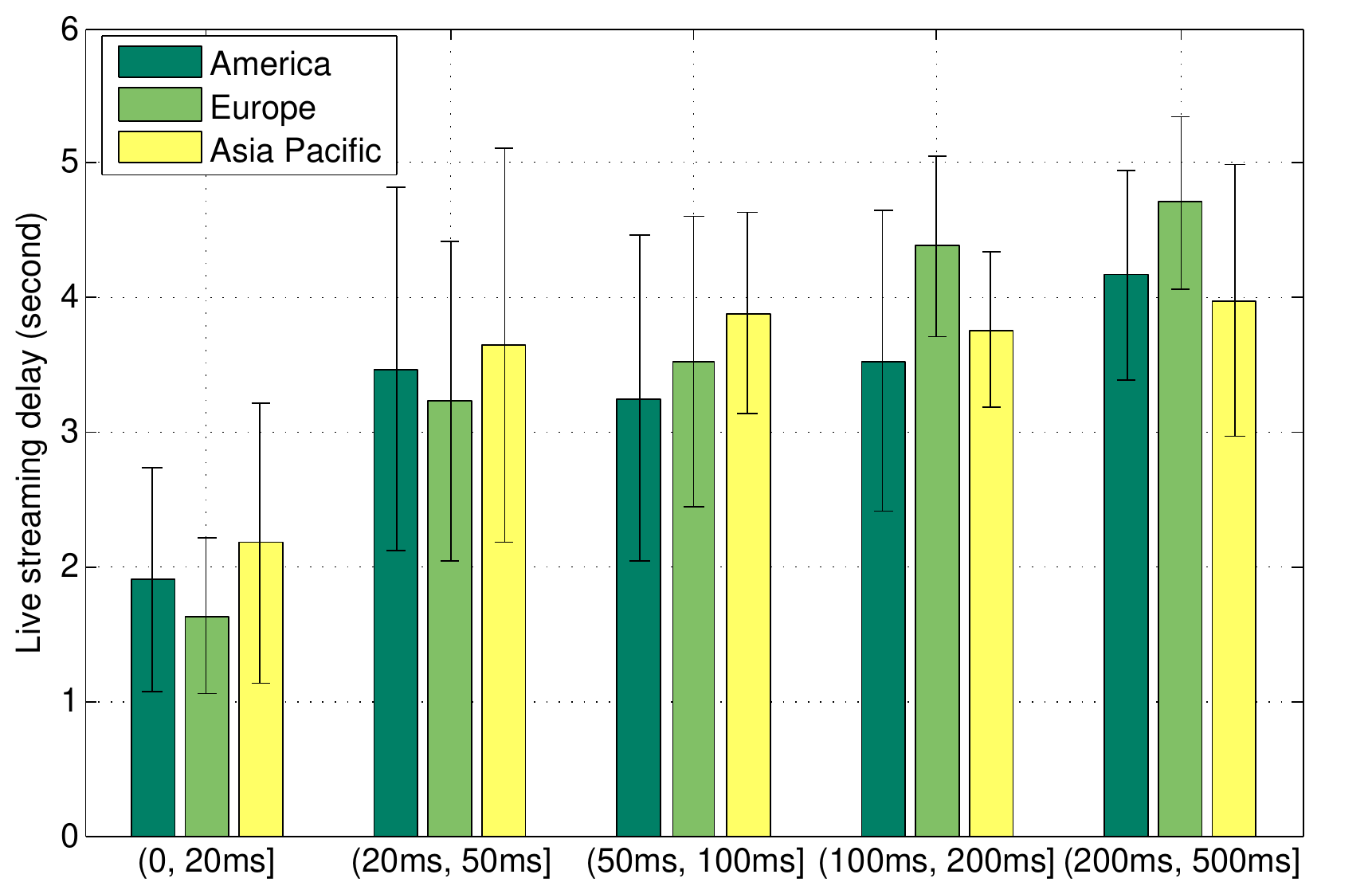}
\caption{Different regions}
\label{gdelay}
\end{minipage}
\begin{minipage}[t]{0.32\linewidth}
\centering
\includegraphics[width=1\linewidth]{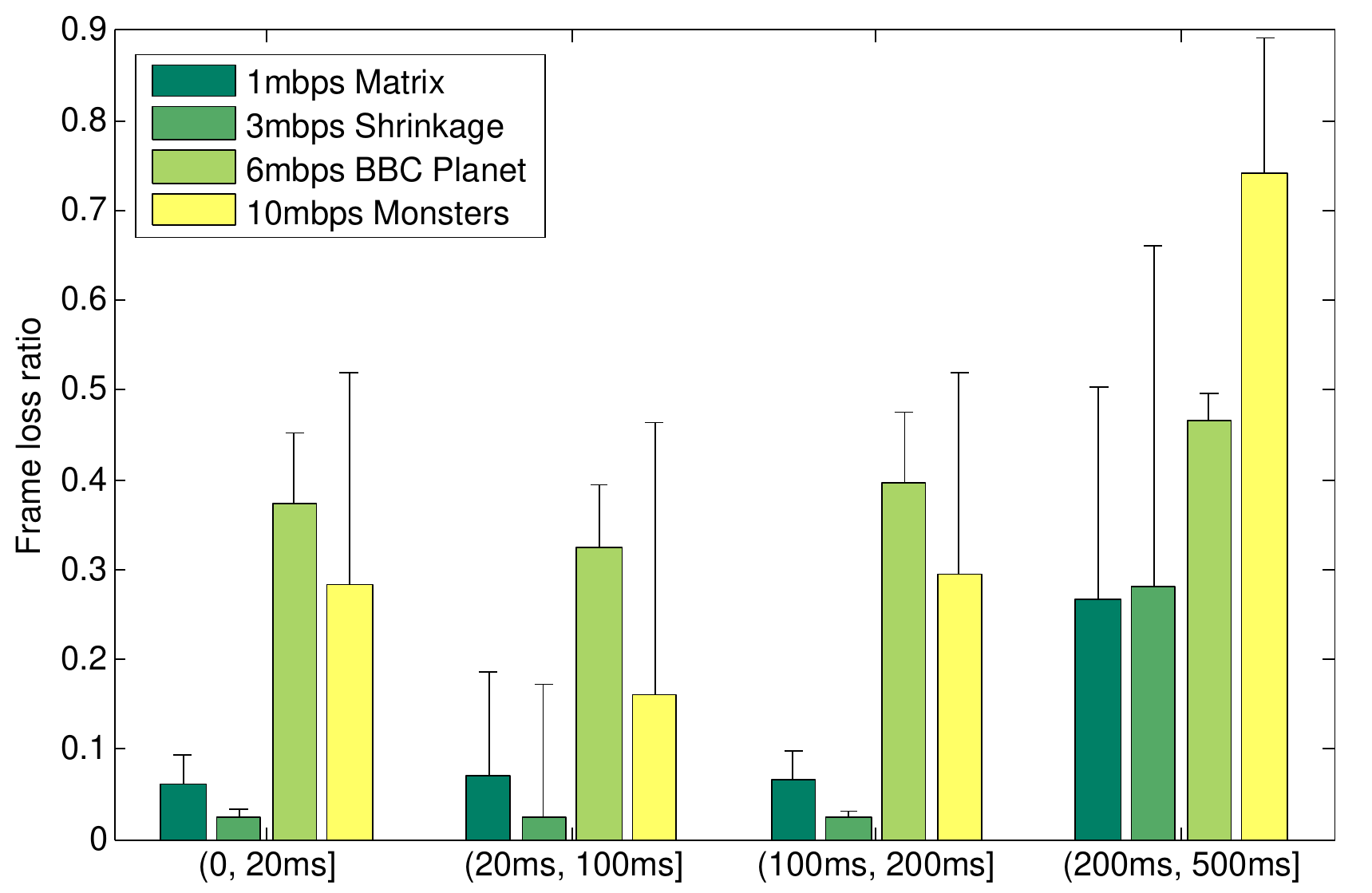}
\caption{Videos with different bitrates}
\label{glost}
\end{minipage}
\end{figure*}

%\begin{figure}[htcp]
%    \begin{center}
%        \subfigure[Different regions]{%
%            \label{gdelay}
%            \includegraphics[width=0.45\linewidth]{globaldelay.pdf}
%        }%
%        \subfigure[Different videos]{%
%            \label{glost}
%            \includegraphics[width=0.45\linewidth]{globallost.pdf}
%        }%
%    \end{center}
%    \caption{%
%       Streaming quality and different RTT latency
%     }
%     \label{fig:global}
%\end{figure}
%%%%%%%%%%%%%%%%%%%%%%%%%%%%%%%%%%%%%%%%%%%%%%%%%%%%%%%%%%%%%%%%%%%%%%%%%%%%%%%%%%%%%%%%%%%%%%%%%%%%%%%%%%%%%
\begin{table*}\small
\tiny
\centering
\caption{\label{Tab:2}Three cloud leasing strategies for crowdsourced live streaming from 7 areas}
\begin{tabular}{|c|c|c|c|c|c|c|c|c|}
\hline
  &  Van (10) &  CA (19)&  VA (20)  &  SA (5) &  K. and J. (20) &  CHN (16)  & S. and A. (4)  & Cost    \\
\hline
TOP preferred first strategy & \multicolumn{2}{c|}{m3 $\times$ 3 (Oregon)} & m3 $\times$ 2 (Virginia)& m1 $\times$ 1 (Sao Paulo)& m3 $\times$ 2 (Tokyo) & m3 $\times$ 1+ m1 $\times$ 1 (Singapore)  &m1 $\times$ 1 (Sydney) & $\$5.584$ per Hour \\
\hline
Centralized provisioning strategy & \multicolumn{4}{c|}{m3 $\times$ 5 (Virginia)} & \multicolumn{3}{c|}{m3 $\times$ 4 (Singapore)} &$\$4.77$ per Hour \\
\hline
Optimal migration & \multicolumn{2}{c|}{m3 $\times$ 3 (Oregon)} & \multicolumn{2}{c|}{m3 $\times$ 2 + m1 $\times$ 1 (Virginia)} & m3 $\times$ 2 (Tokyo)& \multicolumn{2}{c|}{m3 $\times$ 2 (Singapore)}  & $\$5.118$ per Hour \\
\hline
\end{tabular}
\vspace{-0.4cm}
\end{table*}
\section{Performance Evaluation}

We have implemented the crowdsourced live streaming system as a prototype based on PlanetLab, Amazon Cloud, Microsoft Azure Cloud, and the opensource VLC/VLM coder, and have conducted realworld experiments to understand its performance. We have also performed  trace-driven simulations to further evaluate the system performance in large scale.

\subsection{Prototype experimental results}

 In our prototype implementation, both the live crowdsourcers and end users are deployed in 398 Planetlab nodes, which are set up with VLC media player \texttt{0.8.7 Janus} on each node. We deploy the federation of cloud service from Microsoft Azure Cloud and Amazon Cloud in our prototype platform. These two cloud service providers can offer totally $21$ cloud sites distributed all over the world. In each cloud site, the \texttt{General Purpose instances} are provisioned with \texttt{Medium (A2)} from Microsoft Azure Cloud and \texttt{m3.medium} from Amazon Cloud. Each provisioned instance is set up with \texttt{Ubuntu 14.04 LTS} and installed with VLM to manage multiple live streaming channels. Further, we deploy the CloudFront CDN service in \texttt{All Edge Locations} for the globalized content delivery to the geo-distributed viewers. In order to evaluate the streaming quality, the live feeds are generated through videos uploaded from the distributed Planetlab nodes. We use a series of test videos with different resolutions and bitrates \footnote{\url{http://www.cs.sfu.ca/~jcliu/infocom15/crowdsourcing/videos}}. Each dedicated sourcer stores one of these videos as its own live feed. We deploy 18 cloud sites in different regions from Amazon Cloud and Microsoft Azure, 9 from America area, 3 from Europe area, and 6 from Asia Pacific, respectively. To explore the distribution of the 398 planetlab nodes, we measure the RTT latency between the nodes and the cloud sites, and use the cloud site with the minimal latency to approximate their locations. In Fig. \ref{fig:rtt}, we present the nodes population and the average RTT latency from their top 1 preferred cloud sites. With the latency results, each sourcer can construct a preference list of the cloud sites. In order to measure the delay, we implement a live streaming of a timer video \footnote{\url{http://www.cs.sfu.ca/~jcliu/infocom15/crowdsourcing/timer.mkv}} from the planetlab node to the cloud server.  We also use \texttt{ffmpeg} to measure the frame loss ratio during the live streaming through recording the number of duplicated frames (i.e. because the current frame is not received by the playback deadline, the former frame is duplicated) and dropped frames (i.e. the frame is received but corrupted). These planetlab nodes are divided into groups according to the RTT latency in Fig. \ref{fig:rtt}. We present the streaming delay in different areas in Fig. \ref{gdelay}, and frame loss ratio with different videos in Fig. \ref{glost}. Generally, we can see the streaming delay increase more than 80\% if the latency is above 20ms. On the other hand, the frame loss ratio is stable when the latency is under 200ms.

\begin{figure*}[htb]
\begin{minipage}[t]{0.245\linewidth}
\centering
\includegraphics[width=1\textwidth]{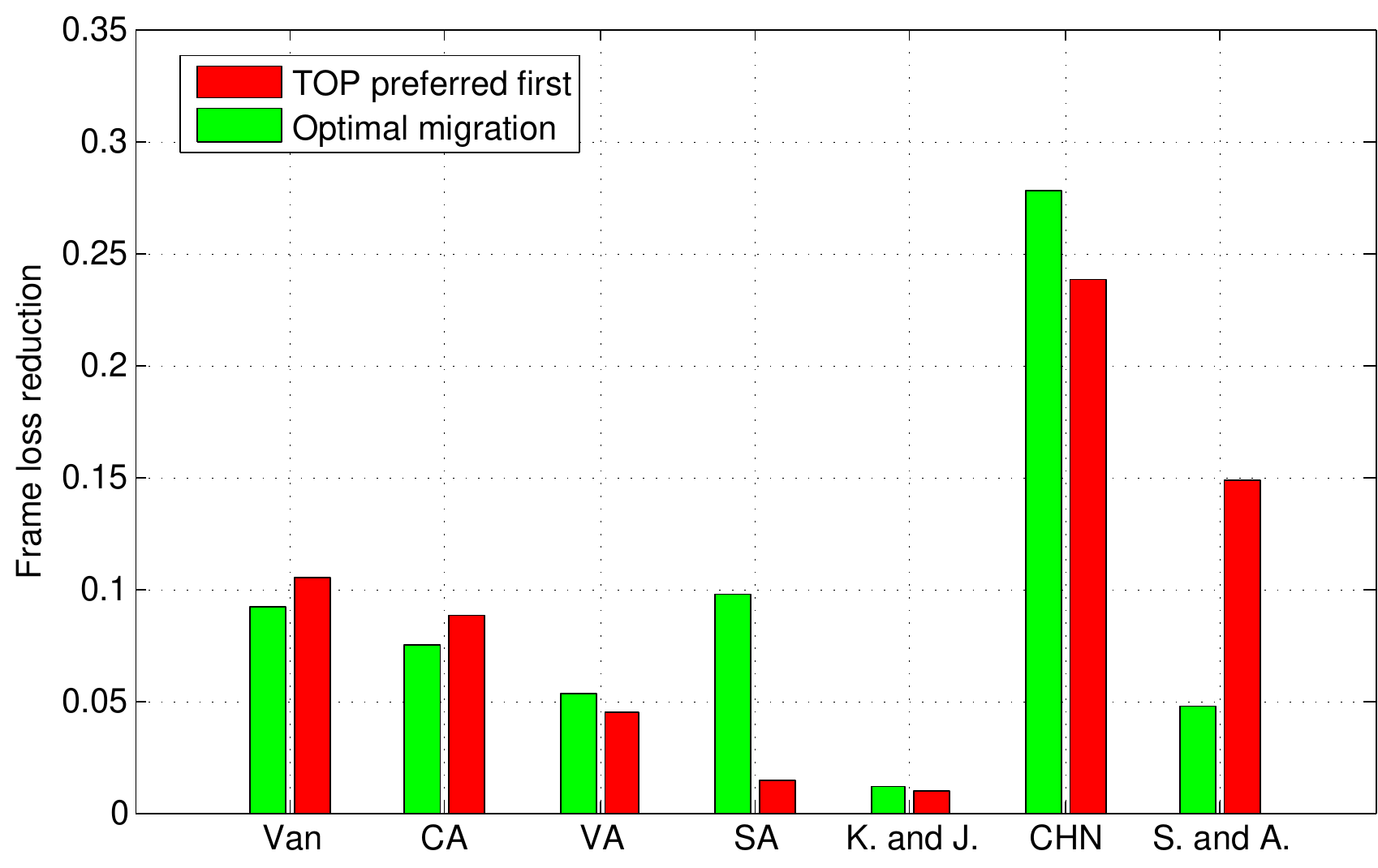}
\caption{Implementation results}
\label{revelant}
\end{minipage}
\begin{minipage}[t]{0.245\linewidth}
\centering
\includegraphics[width=1\textwidth]{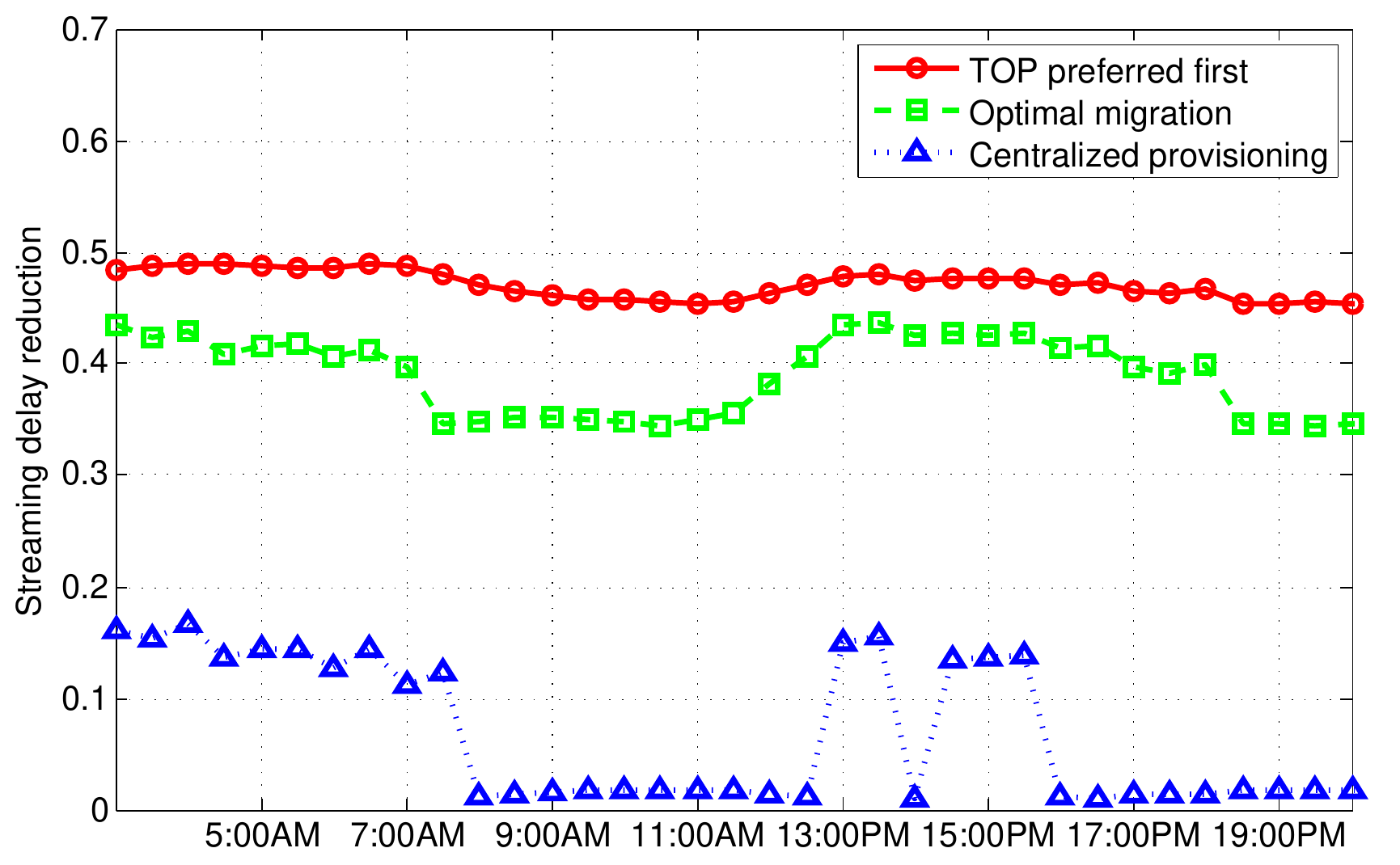}
\caption{Reduction of streaming delay}
\label{reducedelay}
\end{minipage}
\begin{minipage}[t]{0.245\linewidth}
\centering
\includegraphics[width=1\textwidth]{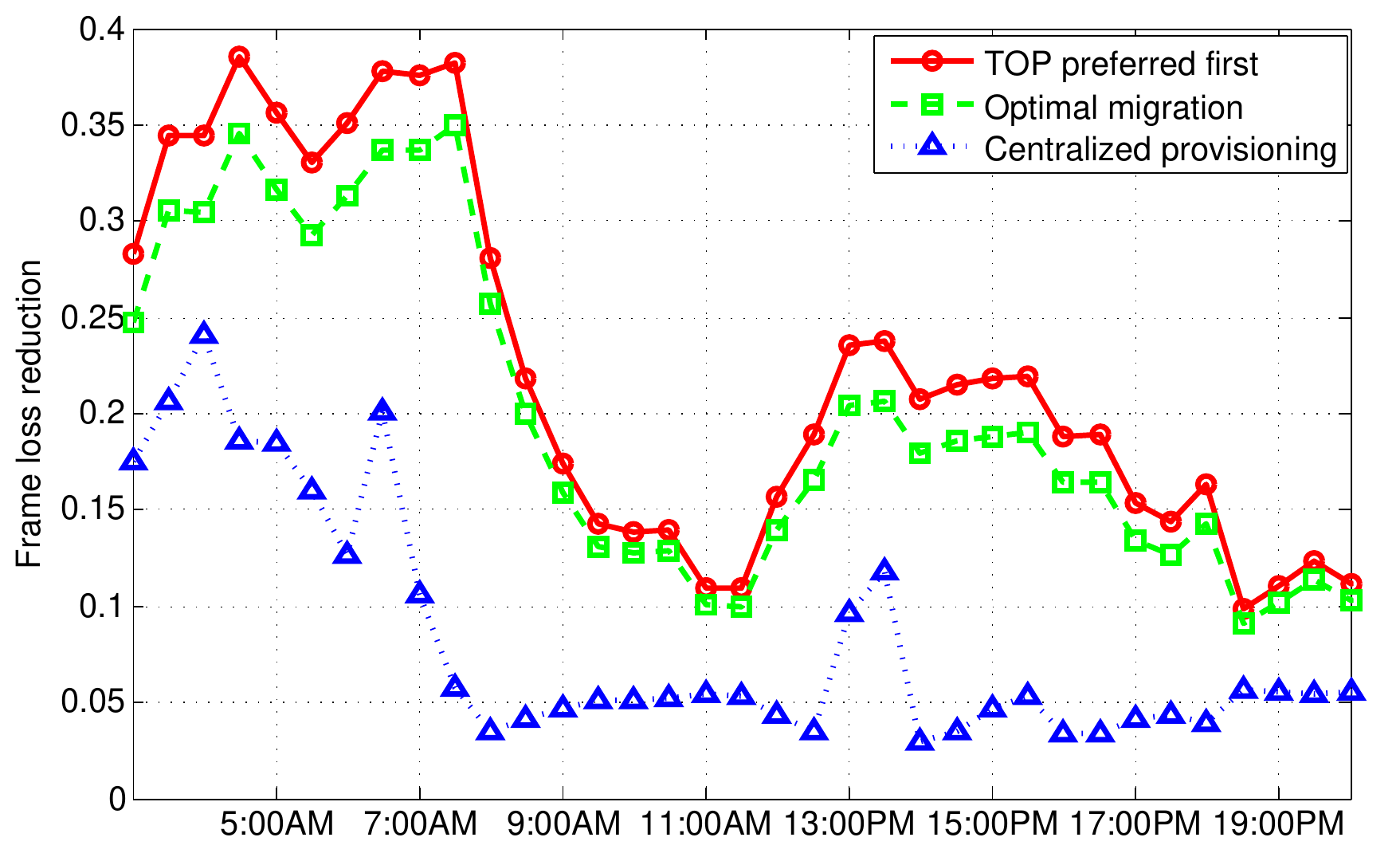}
\caption{Reduction of frame loss}
\label{reducelost}
\end{minipage}
\begin{minipage}[t]{0.245\linewidth}
\centering
\includegraphics[width=1\textwidth]{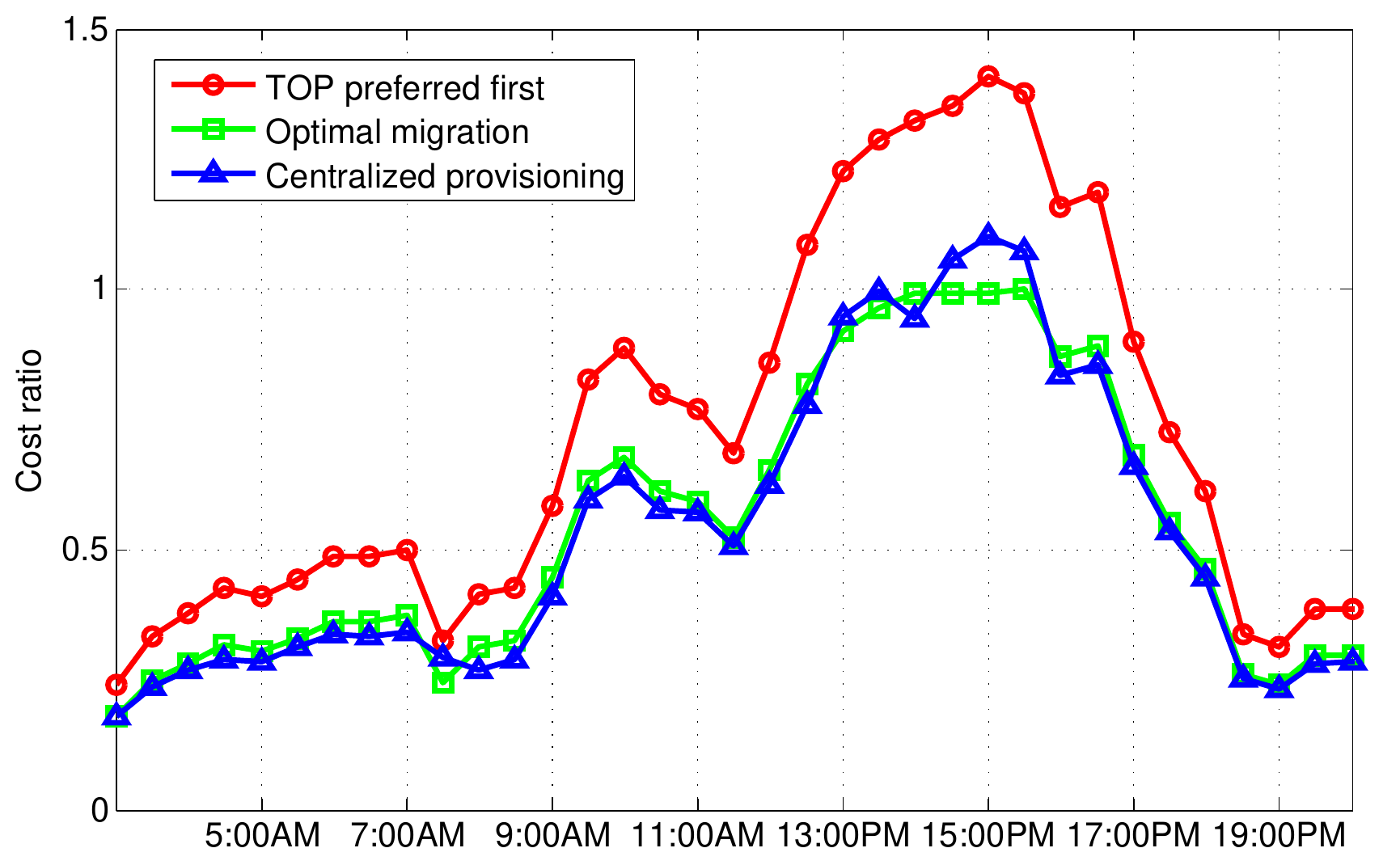}
\caption{Reduction of provisioning cost}
\label{cost}
\end{minipage}
\vspace{-0.4cm}
\end{figure*}

We will further investigate the server provisioning cost and the video streaming quality of the cloud-based strategies through the implementation on the prototype platform. Besides our proposed \texttt{optimal migration} (OM) strategy, two other cloud-based strategies are implemented for comparison. The \texttt{top preferred first} (TOP) strategy deploys all the available cloud sites to allocate the service for sourcers in their most preferred cloud site. Meanwhile, in \texttt{centralized provisioning} (CP) strategy the cloud servers are allocated
in the regions with the most sourcers. Here we select Virginia and Singapore as the central regions, and consider CP as the benchmark strategy. The implementation details of the cloud leasing strategy are presented in Tab. \ref{Tab:2}. For example, m3 $\times$ 1+ m1 $\times$ 1 (Singapore) means one m3.xlarge instance and one m1.large instance are provisioned in Singapore region to serve 16 sourcers. We also calculate the server provisioning cost per hour according to the prices of Amazon EC2. CloudFront is deployed as CDN for the global distribution, and we record the average frame loss ratio from 20 distributed users. Generally the frame loss ratios can be reduced by about 10\% for TOP and OM strategies. Especially, for the plantlab nodes in China, the improvement can reach almost 30\% with the proposed strategy. Comparing with TOP strategy, our proposed solution saves 8.34\% cost, and improves 9.1\% video quality on average.

%%%%%%%%%%%%%%%%%%%%%%%%%%%%%%%%%%%%%%%%%%%%%%%%%%%%%%%%%%%%%%%%

\subsection{Trace-driven simulation results}

To further evaluate the performance of the proposed strategy in larger
scale, we simulate the system with the real world trace data from
Twitch.tv and the measurement results from the prototype system. The diverse prices of distributed cloud sites are referred to Amazon Cloud and Microsoft Azure Cloud. We
consider a conventional \texttt{centralized dedicated server} (CDS) strategy as the
benchmark, in which the single server is allocated in the central
region to service the global requests. The price cost should cover the
peak user demand, and we will take this cost as the budget constraint
in our proposed OM strategy. We also set $p/q=0.1$ and the preference
value is inversely proportional to the RTT latency. Another
two cloud based strategies are deployed for comparison.  All these cloud-based strategies can scale their provisioning capacity adaptively to the user
demand.

Fig. \ref{reducedelay} shows the streaming delay reduction of the three cloud-based strategies comparing with the benchmark CDS strategy.
Generally, TOP and OM strategies, which deploy the geo-distributed cloud service, can reduce almost $50\%$
streaming delay of the benchmark strategy. The CP strategy can have an improvement only when most of viewers concentrate on several sourcers from
the same region (e.g. 3:00AM-8:00AM in Asia and 13:00PM-16:00PM in Europe). Different from the streaming delay reduction, the frame loss reduction is more dynamic with time
variations in Fig. \ref{reducelost}. Before 8:00 AM, most of popular sourcers are from Europe and Asia, the CDS strategy would suffer from the long
transmission, despite the total number of streams is not large, and there is still extra available bandwidth capacity for the rented server. After 9:00AM, sourcers from north
America attract more viewer demand. Then dedicated server can provide an acceptable service with less frame loss ratio.
In Fig. \ref{cost}, we present the cost ratio between the three cloud-based strategies and the benchmark strategy. As the server instances are allocated in the distributed cloud sites with diverse prices, the Top1 strategy can lead to a higher cost when the peak demand comes.
Because of the budget constraint, the provisioning cost in our proposed strategy is limited under the cost of the benchmark. Yet, comparing with the TOP strategy, the gap of streaming delay and frame loss ratio can still be kept within $5\%$, and almost $30\%$ of the provisioning cost is saved through the service migration during peak demand.

%\subsection{Implementation on Prototype Platform}

\section{Conclusion and Future Work}
In this paper, we explored the emerging crowdsourced live streaming systems, in which both the number and distribution of the crowdsourcers can be highly dynamic. It further motivated the design of cloud leasing strategy to optimize the cloud site allocation for geo-distributed live crowdsourcers. A prototype of crowdsourced live streaming platform was built with Amazon Cloud/Microsoft Azure and Planetlab nodes. The performance of the proposed strategy was evaluated through extensive experiments.

 Our work is an initial study, and there are still many open issues to be further explored. We plan to continue enhancing our design by conducting more evaluations on our prototype with larger scale experiments. Our ongoing work includes tailoring our method for some specific crowdsourced live streaming applications, such as synchronizing multiple collaborative crowdsourced live videos for 3D immersive environment reconstruction or real-time interaction. We are also interested in extending our current deployment strategy to a more general scenario, in which the distributed server instances can cooperate with CDNs for a larger service coverage with a lower cost. In addition, we believe that the dynamic geo-distributed crowdsourcers are predictable, in which there are two major types of live sources, namely, scheduled sources and non-scheduled sources. The scheduled sources mean the crowdsourcers follow some social event during a certain time, such as a presidential election, or a football match, which is easy to predict. As to the non-scheduled sources, the crowdsourcers can start their live streaming arbitrarily. The time-varying live sources usually relate to the dynamic viewers demand, since the crowdsourcers are motivated to get more subscribers as a reward. They tend to broadcast in a fixed time every day, or choose a period when a peak number of viewers can be achieved. This behavior of crowdsourcers is evident in some modern crowdsourced live streaming platform, such as Twitch.tv. Our solution could be enhanced with crowdsourcer prediction through user behavior analysis from real-world measurement results.


\section*{Acknowledgment}
This research is supported by Natural Sciences and Engineering Research Council of Canada (NSERC), a Start-up Grant from the University of Mississippi, a Start-up Grant from the Jiangnan University, the National Natural Science Foundation of China (No. 61103223) and the Natural Science Foundation of Jiangsu Province (No. BK2011003).



%\bibliographystyle{IEEEtran}

\end{document}